# Identity theft and societal acceptability of electronic identity in Europe and in the United States

Marek Tiits*, Tarmo Kalvet, David McBee

## Abstract

This paper addresses critical questions surrounding the security of government-issued identity documents and their potential misuse, with an emphasis on understanding the perspectives of ordinary citizens across Europe and the United States of America. Drawing upon research on technology acceptance and diffusion, the research focuses on understanding the factors that influence users' adoption of novel identity management solutions. Our methodology includes a comprehensive, census-representative survey spanning citizens from France, Germany, Italy, Spain, the United Kingdom, and the USA. The paper's findings underscore a robust confidence in government-issued identity documents, contrasted by a lower trust in private sector services, including social media platforms and email accounts. The adoption of artificial intelligence for identity verification remains contested, with a significant percentage of respondents undecided, indicating a need for explicit explanation and transparency about its implementation and related risks. Public sentiment leans towards acceptance of government data collection for identification purposes; however, the sharing of this data with private entities elicits more apprehension.

## 1. Introduction

As societies continue to transition towards the digital ones and with more and more facets of our daily lives intertwined with digital platforms, from finance to healthcare, the reliability and security of our identity markers become indispensable anchors in this digital realm. Identity documents, traditionally used as authoritative assertions of one's identity, are now part of a critical infrastructure, underscoring transactions of trust and authentication across numerous digital services (e.g., Kemppainen et al. 2023).

However, with the development of digital technology, the ability to forge or manipulate data – including biometrics technology and its realism – develops as well. With the increasing availability of different tools, digital forgery has become massive and widespread. As Boneh et al. (2019) have argued, "the barrier to entry for manipulating content has been lowering for centuries. Progress in machine learning is simply accelerating the process". Today, there are hundreds of different technologies and programmes available to forge or manipulate identity data, or otherwise engage in identity theft. These can be spoofing attacks, adversarial attacks or digital manipulation attacks (Dang et al. 2020). Identity theft has become a significant concern for individuals, organizations, and businesses, and has directed all relevant stakeholders to work on secure digital identity solutions. Accordingly, digital security has become the cornerstone for further development of the information society.

---

* Institute of Baltic Studies, Tartu, Estonia, marek@ibs.ee

Obtaining someone else's personal information or identity document, such as an identity card or passport, is where identity fraud begins, and it is becoming increasingly widespread (Reyns 2018, Akdemir 2021). With a stolen identity, the fraudsters can effectively become someone else, allowing them to access the victim's financial or other accounts, access communications, set up new contracts, or present false information to the authorities. This is not only a violation of privacy but may bring about substantial financial and/or legal consequences to the victim. Evidence is also available on the associated major social and psychological impacts (Betz-Hamilton 2022, Akdemir 2021, Kalvet et al. 2019a).

Securing the authenticity of identity documents is key to ensuring the continuity of identity, dependability of identity checks and forensic investigations. However, the growing complexity of document security and border control pressures often make it impossible to confirm whether documents are genuine or counterfeited. Building on technology acceptance and diffusion research, this article focuses on understanding which factors influence users' adoption of novel identity management solutions.

Fighting identity crime and the establishment of a more secure identity should start from sufficient insights into identity crime, i.e. how people fall victims of identity theft. Understanding which novel identity management solutions they are more likely to accept allows, in turn, for the establishment of a more secure identity. This is why the current research sets to analyse the misuse of identity in a number of major European countries, namely in France (FR), Germany (DE), Italy (IT), Spain (ES), United Kingdom (UK), and for comparative purposes also in the United States of America (US). Furthermore, we test the societal acceptability of a number of identity management solutions. We intend to identify, among other topics, if and how does the personal experience with identity theft explain the perceptions on the use of biometric data and advanced data processing solutions in identity management. For informing the current research, a census representative on-line survey was carried out in native languages in the above-mentioned countries in March-April 2022. From each of the seven countries at least 500 fully completed survey questionnaires were collected.

In the following chapter, "Literature" (Chapter 2), earlier studies and the key academic literature on societal acceptability of technology are discussed. Thereafter, analytical framework and the means of data collection are briefly presented (Chapter 3), and the results on common identity document use patterns, identity theft and the societal acceptability of modern identity solutions are presented (Chapter 4). The current paper concludes with recommendations that facilitate the adoption of modern identity solutions (Chapter 5).

## 2. Literature
### 2.1. Technology adoption and diffusion

Technology acceptance continues to be an important research issue in social sciences and the development of respective models has received increasing attention in the recent decades. The general point of departure is that there are several factors that will influence the user as to whether or not they adopt technology. The goals of many studies have been to find factors that can be used to motivate the user to accept and start using the new technology (Ash 1997, Mathieson 1991, Venkatesh 2000).

In the course of the research numerous frameworks and models that have been conceived to study the user adoption of emergent technologies and researching the factors that influence this acceptance, were mapped. These include the Technology Acceptance Model (Davis 1989), Theory of Planned Behavior (Ajzen 1985), Diffusion of Innovation theory (Rogers 2003), the Theory of Reasoned Action (Fishbein and Ajzen 1975), to name some. These and other similar frameworks form the backbone of numerous studies on technology acceptance, while other research combines previous models or introduces new constructs to expand upon the existing frameworks (Taherdoost 2018).

One popular model that considers relevant factors in technology adoption is the Technology Acceptance Model (TAM), which suggests that perceived usefulness and ease of use play a significant role in the adoption of a technology (Davis 1989). It is widely used to understand and explain the use of information technologies in different contexts (King and He 2006). Another model, the Unified Theory of Acceptance and Use of Technology (UTAUT), is more comprehensive and includes additional factors. It is currently one of the most widely used technology acceptance models (Venkatesh et al. 2003). According to UTAUT, four key constructs influence technology adoption: performance expectancy, effort expectancy, social influence, and facilitating conditions. Additionally, variables such as gender, age, prior experience with related technologies, and voluntariness of use are considered to influence the intention to use and actual use behaviour. There are further elaborations of UTAUT, such as UTAUT2 (Venkatesh et al. 2012) and an extension of UTAUT2 (Tamilmani et al. 2021), but the foundations remain largely the same.

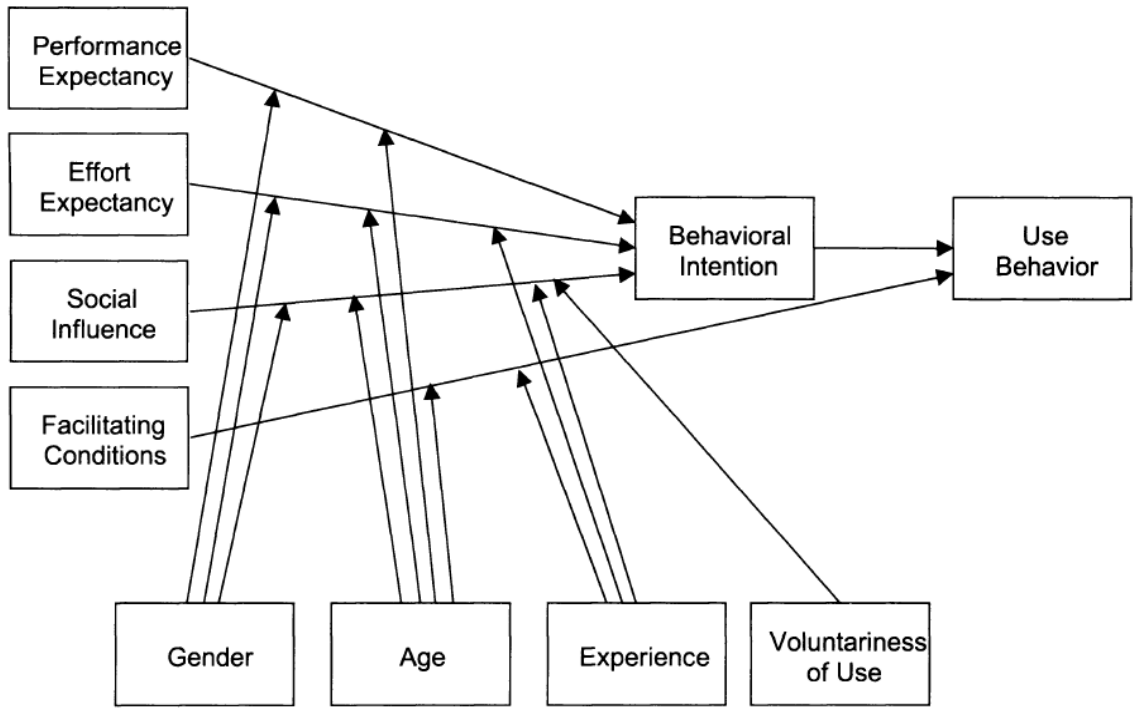

*Figure 1. Schematic view of the Unified theory of acceptance and use of technology*

Source: Venkatesh & Davis 2003, p. 447.

As UTAUT is a commonly embraced model for examining the acceptance of digital technologies, this theory served as the foundational structure for the authors in creating a framework for empirical data collection. Applying this model to the analysis of the social acceptability of modern identity solutions, based on the literature review, the following interpretations were made:

- Performance expectancy encompasses expectations related to the adoption and use of identity solutions. This includes direct benefits such as increased security of government-issued identity solutions and private sector accounts, as well as protection against identity misuse. Performance expectancy also considers social risks associated with technology, such as privacy protection, oversight of artificial intelligence-driven decisions, and government data sharing. It is crucial to address these concerns, as the perception of reduced benefits due to function creep may decrease the intention to use and actual adoption.

- Effort expectancy is associated with the ease of using a technology. While physical identity documents are generally easy to use, electronic identity solutions like two-step authentication or electronic identity wallets may require specific equipment or advanced technical knowledge.

- Social influence refers to the influence of friends, family, or other individuals, including role models and opinion leaders, who can either encourage or discourage the use of modern identity solutions. This influence is particularly relevant for "less informed" individuals who may rely more on word-of-mouth recommendations rather than formal communication channels like newspapers or government documents.

- Facilitating conditions, according to the UTAUT model, also play a role in technology adoption. For instance, familiarity with modern information and communication technology (ICT) is necessary to use electronic functions like automated border control gates. However, in Europe, there is widespread availability of such skills, as demonstrated by high levels of personal computer and internet usage (Bilbao-Osorio et al. 2014; Eurostat 2014).

Furthermore, variables such as gender, age, experience with specific or related technologies, and voluntariness of use are considered influential in the adoption process (Venkatesh et al. 2003). Thus, demographic variables like gender, age, education, and occupation are important for the current study, as they help identify different social groups. For example, younger generations, who are generally more technologically savvy and eager to adopt various ICT solutions, may possess better knowledge about securing online accounts and have more informed opinions on advanced technology topics. Finally, the voluntariness of the use of technology is another factor highlighted in the UTAUT model.

When developing survey questions for the study, both the technology acceptance and diffusion research and the findings of earlier studies were taken into account.

## 2.2. Empirical research on identity theft and the acceptability of technology

Our earlier research (Tiits and Ubakivi-Hadachi 2016, Kalvet et al. 2019a) has concluded that roughly 25-30% of the population of Austria, France, Germany, Italy, Spain, United Kingdom have experienced some form of attempted or confirmed misuse of personal information over the period of 2013-2015. Only 10% of these cases were detected before personal information was actually taken. Thus, around 100 million citizens were forced to take extra steps to protect their identity during a 3-year period in the EU. Almost half of them had to do so more than once, as they experienced multiple incidents. As a result of the misuse of personal information, close to 40 million EU citizens have experienced significant personal consequences, such as debt collectors contacting them, problems with their family or friends, being denied a new service, having to face legal problems, etc.

The total value of the money, goods or services obtained by criminals from 2013 to 2015 was roughly 12-16 billion euros in the EU. This is, however, only the "consumer side". From the misuse of personal information, various institutional actors, e.g., financial or health insurance institutions, are likely to have incurred additional financial losses that are unknown to the individuals and, therefore, not reflected in this study (Tiits and Ubakivi-Hadachi 2016, Kalvet et al. 2019a). For instance, in the United States of America, Internal Revenue Service has estimated that it paid 4 billion euros in fraudulent identity theft refunds in filing season 2013, while preventing fraudulent refunds of 18 billion euros (based on what they could detect) (U.S. Government Accountability Office 2014). It is within reason than to assume that, given the above examples, the rough financial cost of identify in Europe reflects only the tip of the iceberg.

Other studies such as those commissioned and co-operated on by the United States Department of Justice and the Bureau of Justice Statistics (Harrell 2014, Oudekerk et al. 2018) have studied identity theft issues in recent years and confirm the scale and growth of the problem. Javelin's 2020 Identity Fraud Study concludes that total identity fraud reached 15 billion euros in 2019 while criminals are targeting smaller numbers of victims and inflicting damage that is more complex to prevent or remediate. The research states that "the type of identity fraud has drastically changed from counterfeiting credit cards to the high-impact identity fraud of checking and savings account takeover. At a time when consumers are feeling financial stress from the global health and economic crisis, account takeover fraud and scams will increase" (Javelin 2020).

Eurobarometer survey on cyber security from 2020 (Kantar 2020) is also reflecting raising concerns: as compared to the study from 2017 (TNS Opinion & Social 2017), less Europeans feel they can protect themselves sufficiently against cybercrimes (59%, down from 71% in 2017). Three key concerns are related to falling victim to the bank card or online banking fraud (67%), the infection of devices with malicious software or identity theft (both 66%), and 6% of the respondents have actually experienced identity theft 2017-2019 (Kantar 2020). As an example, Apple Inc. reported recently that its App Store security features blocked nearly 3.9 million stolen credit cards from being used. In total, App Store prevented over 1.9 billion euros in fraudulent transactions in 2022. (Apple 2023)

Earlier research has shown that the public has little trust in the security of popular Internet services, such as e-mail or Facebook (Tiits and Ubakivi-Hadachi 2015). Widespread phishing attacks and misuse of Internet accounts, bank accounts and credit cards do not foster trust in these services. Nonetheless, confidence in government issued electronic identity cards and passports remains very high (Tiits et al. 2014a, Tiits et al. 2014b) and is likely to be because the misuse of government issued identity documents remains infrequent in citizens' view as compared to other forms of identity fraud.

Also, the importance and acceptance of biometric solutions have increased considerably, and such technologies should be preferred in identification solutions. The direct aim of biometric technology (which includes biometric identifiers like face and fingerprints) is to enhance the reliability of identification. Biometrics is a tool used to identify and reliably confirm an individual's identity based on physiological or behavioural characteristics (or a combination of both) that are unique to a specific human being. Since biometrics provides a close link between the physical person and identity credential, e.g., a government issued identity document, it is considered a strong form of identification technology (Tiits et al. 2014b).

Likewise, Zhong and colleagues (2021) investigate the possible factors that drive customers' willingness to utilise facial recognition payment; the findings showed that factors such as perceived enjoyment, facilitating conditions, personal innovativeness, coupon availability, perceived ease of use, perceived usefulness, and users' attitude are main drivers of customers' decisions to use facial recognition payment.

In sum, the adoption of biometrics or electronic identity solutions more broadly is always about balance between 'benefits' and mitigation of societal 'costs' / 'risks'. Although impressive progress has been reported in the fields of biometrics, numerous technological, legal and social acceptance related challenges remain (Tolosana et al. 2022).

# 3. Research framework and data collection

## 3.1. Research questions

In the following, we delve into the research questions that guide our investigation on identity management solutions and the perception of citizens in Europe and the United States. Building upon the existing body of knowledge, we aim to expand our understanding of the threat of identity theft, explore the willingness of individuals to adopt emerging technologies such as eID and biometrics, and examine the influence of socio-economic background and risk perception on the acceptance of modern ID solutions. By addressing these research questions, we seek to contribute to the existing literature and provide insights into the evolving landscape of identity management in these regions.

The current paper goes beyond the state of the art by analysing the public perception on identity documents in different countries, and especially in the context of identity theft, i.e. someone obtaining or using someone else's personal information without his or her permission, to pretend to be the person in question or to carry out business, or engage in other types of activities and transactions in that person's name without their permission. To inform the technology choices, we are, in this context, interested in the willingness of the public to opt in the future for more secure identity management.

The more specific research questions that we address are the following:

RQ1: How confident is the general public in the Europe and in the United States in security of government-issued identity solutions (e.g., passports, identity cards) and private sector accounts (e.g., financial accounts or social media accounts)?

RQ2: What is the extent of the identity theft threat in Europe and the United States, and which identity solutions are at greater risk?

RQ3: To what extent are citizens in Europe and the United States open to adopting modern identity solutions, and how to increase societal acceptability when designing the procedures for issuance and renewal of identity documents, identity checks and verification of identity documents, and government data sharing?

While addressing the aforementioned research questions, we also examine whether individuals who have personally experienced (an attempted) identity theft are more inclined to support the implementation of contemporary identity solutions. Additionally, we explore whether socio-economic factors, including education level and occupation, as well as individuals' perception of privacy risks, influence the general public's willingness to accept modern identity solutions.

## 3.2. Data collection

The current research builds on a unique new data set that allows us to analyse the identity theft, and the public perception on the novel identity management solutions, in five European countries and in the United States of America, in a comprehensive manner. The research also addresses the weakness of previous national studies - the results of some of the earlier (national) studies are not comparable either, as the definition of identity theft and the exact wording of survey questions varies from one study to another. Also, the strength of the current survey is the focus on technology scenarios, to conclude with recommendations that facilitate the adoption of modern identity management solutions.

A survey on the identity theft and societal acceptability of modern identity solutions was carried out from 23 March to 8 April 2022 in the following six countries: France (FR), Germany (DE), Italy (IT), Spain (ES), United Kingdom (UK) and the United States of America (US). These countries represent a selection of countries in Europe, plus, for comparative purposes, North America.

The above-mentioned countries represent technologically advanced countries with different socio-economic and cultural contexts. Furthermore, establishment of identity and identity management are also handled differently in different countries covered by this study. Some governments (e.g., Spain) are relatively more advanced in issuing electronic identity cards, while there are also governments (e.g., the United States) that do not issue any identity cards at all. Some countries embrace the idea of the establishment of a unified national database on all issued identity documents, while others continue to be hesitant on this. Some of the countries are already experimenting actively with electronic identity wallets, while the others are still contemplating the idea.

For the data collection purposes, a comprehensive survey questionnaire with 52 questions was developed (Tiits & Kalvet 2022). The survey questionnaire was designed in such a way that a substantial part was shown only to the respondents who have personally experienced identity theft within the last 36 months. Accordingly, a major share of the respondents filled in a shorter version of the questionnaire. The survey was carried out as an online survey using Alchemer (Alchemer 2023) web survey platform. Cint web survey panels (Cint 2023) were used to target and recruit individuals between the ages of 18 and 64 from pre-enrolled respondent databases. The collected responses are representative of the gender, age and regional distribution of the population of respective countries.

At least 500 complete responses were collected from each of the six countries covered by this survey. When carrying out the survey, age, gender and regional balance that represents the demography of the respective countries was followed. The responses that showed obvious signs of the lack of responder's attention or had been filled in unrealistically speedy manner, in less than 180 seconds, were removed from the survey data set. As the result, 2,950 fully completed responses were retained for analysis. The age breakdown for the respondents is presented in Figure 2.

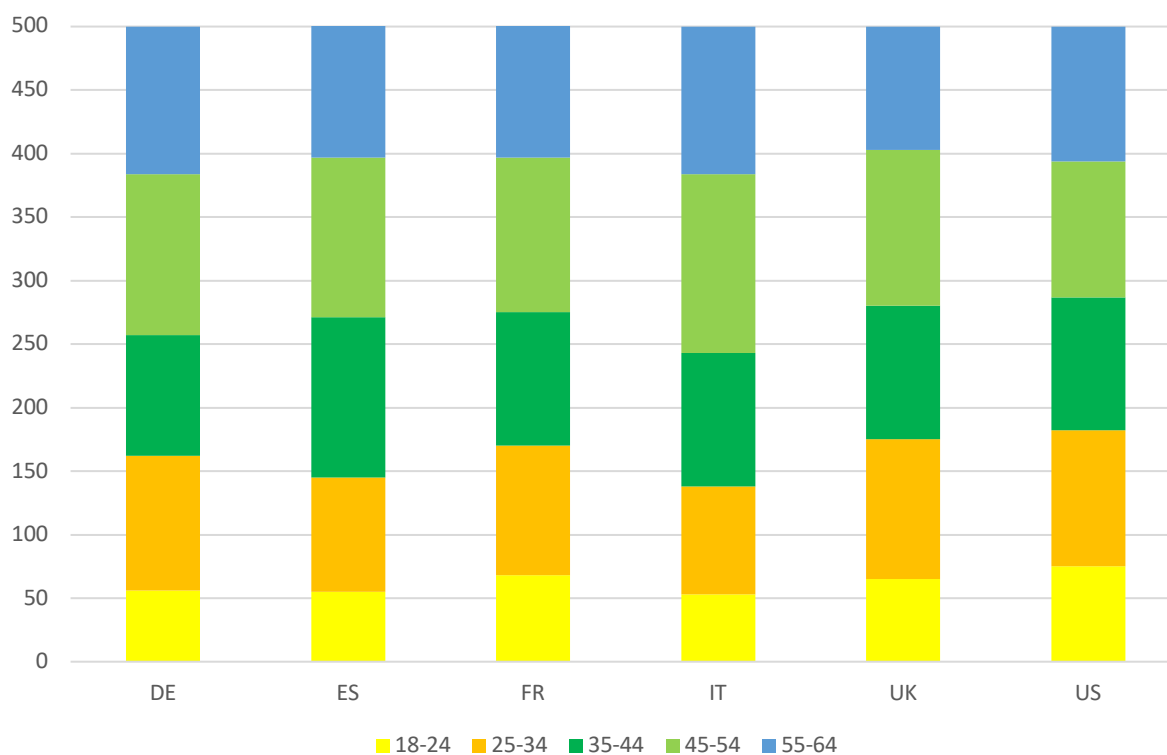

Figure 2 – Age breakdown of the survey respondents

The above data collection comes with an obvious limitation, though. According to Eurostat (2022), approximately 10% of the 16-74 years old population do not use the Internet and are thus automatically excluded from online surveys. We are aware of this inherent weakness of the survey data set and acknowledge in the subsequent analysis that online surveys exclude a minority who has no sufficient knowledge or skills for using Internet. However, such people are generally of older age (Brandtz et al. 2011) many of whom would be less likely to be able respond to the questions that deal with advanced technologies. Thus, we expect that the share of persons who are uninformed or undecided about various technology specific questions would have been greater if we would have been able to cover also persons not using Internet.

Statistical tools utilised in this study include cross tabulations, analysis of variance, chi-square tests, simple and multiple linear regression, as well as logistic regression models. Thus, data distribution, relationships between variables or sets of variables, as well as possible strengths and directions of these relationships are explored. Both effect sizes and the significance of results were accounted for. Every variable was carefully explored, cases were checked, and outliers were excluded from the subsequent analysis. Statistical tests are carried out with the Stata software package. To do so, a comprehensive dataset was created and cleaned in Stata IC 15.1.

## 4. Results
### 4.1. Confidence in government-issued and private sector identity

To gauge public confidence in government-issued identity documents and private sector accounts, we asked respondents to report their confidence that various types of documents and accounts were secure, in terms of their design and use. Overall, respondents are confident about the security of their government issued identity solutions. Respondents consider even birth certificates safe (88%), even though they do come in various forms and shapes in different countries, and have, typically, virtually no modern security features. (See Figure 3, below.)

Interestingly, respondents whose birth certificates, passports, ID cards, licenses, residence permits, or social security cards, were misused did not experienced lower levels of confidence in those same documents. A series of ordered logistic regressions with misuse and confidence failed to yield significant influences (p > .05) for birth certificates, passports, ID cards, licenses, residence permits, and social security cards. (These results are available upon request.)

*Figure 3 – How confident are you that your current identity documents, which the government has issued to you, are secure both in terms of their design and how they are actually issued?*

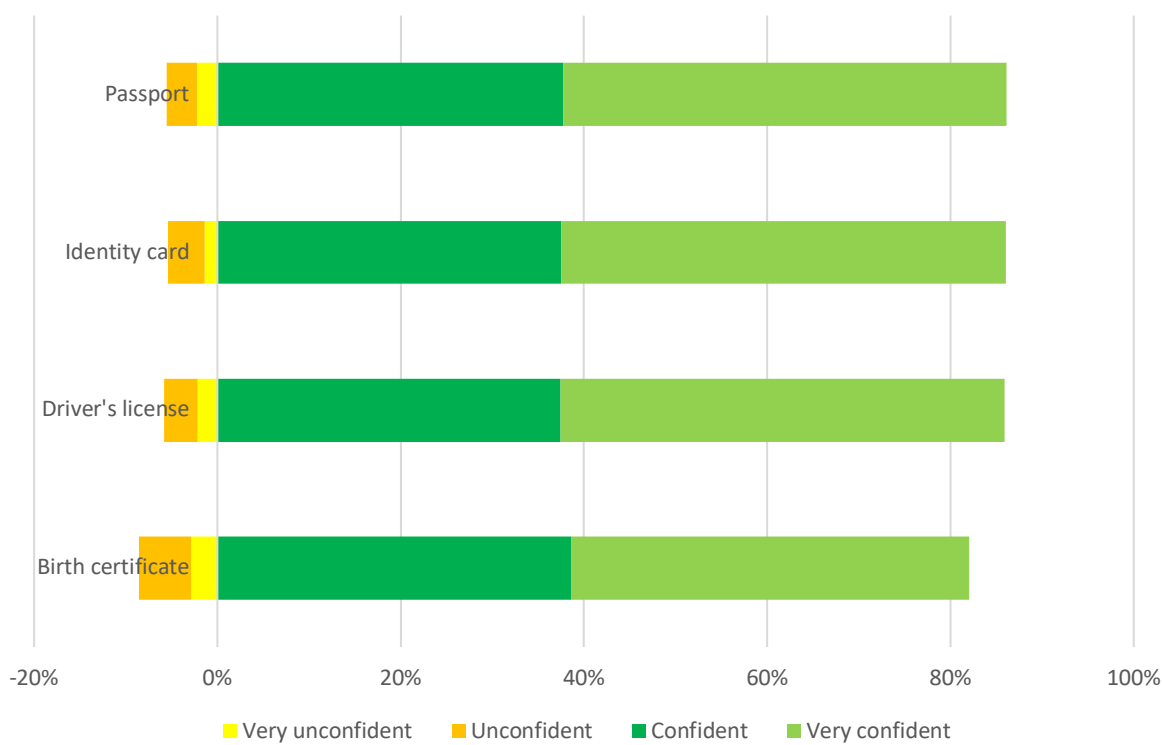

*Note: Undecided respondents are excluded from this figure. The sum is therefore less than 100%.*

The everyday use of government issued identity documents varies significantly across countries, subject to availability of the identity card. 85-95% of respondents consider identity card to be their primary identity document in DE, ES, FR and IT. Driver's licenses are a dominant document in the UK (48%) and in the US (61%). When choosing, which identity document to carry, the public tends to prefer the document that is more convenient to carry daily or has a broader set of uses. The avoidance of a potential misuse, e.g., when the document is lost or stolen, is not, for the majority of respondents, a primary consideration. 50% of the respondents, who have more than one identity document, prefer a specific document because it is more convenient to carry on daily basis because of its (compact) format; and 35% of the respondents mention a broader set of uses. Only 12% of the respondents like to carry a particular document because it is more difficult to misuse, e.g., when lost or stolen. Respondents are confident that their bank accounts and credit cards are secure. However, there is a lot of scepticism about security of social media accounts, such as Twitter or Facebook. (Figure 4)

*Figure 4 – How confident are you that the following services, which you have yourself signed up for, are provided in a secure manner so that you are protected from any potential misuses?*

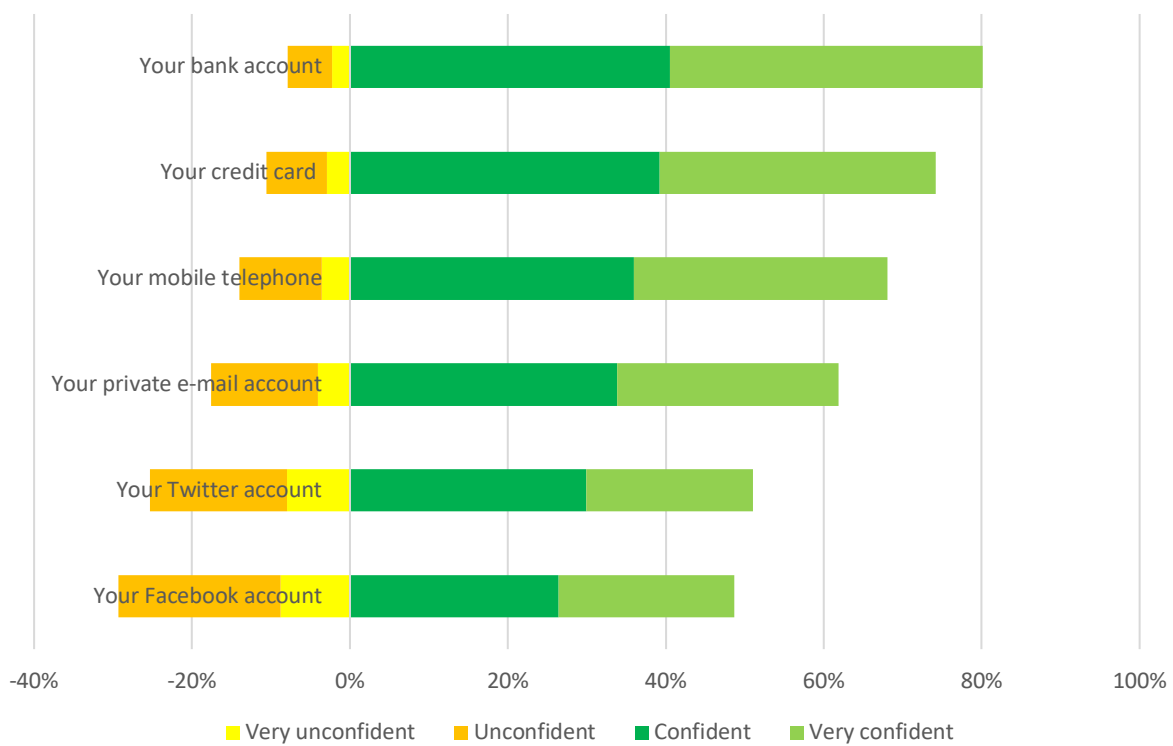

*Note: Undecided respondents are excluded from this figure. The sum is therefore less than 100%.*

Generally, respondents are more confident about government-issued identity solutions than private sector accounts. Notably, the lowest level of confidence of any government-issued identity document (83% for birth certificates) is higher than the highest level of confidence for any private sector account (81% for bank accounts). A mean-comparison test with unequal variances shows that this difference is statistically significant (p = .006).

## 4.2. Identity theft in Europe and in the United States
### 4.2.1. Personal experience with identity theft

To map the extent of identity theft in Europe and the United States, we asked respondents to indicate if anyone had misused or attempted to misuse one of their identify documents or electronic identity solutions in the past 36 months without their permission. Across the countries, 16% of the respondents indicated that someone had (attempted) to misuse their identity. However, this varies by country. Whereas 30% of respondents from the United States indicated having experienced such misuse, only 9% of their United Kingdom counterparts reported similarly[1]. From such misuses and attempted misuses, 33% involved identity cards, 28% mobile ID, 27% passports in 27%, and 23% driver's licenses. (Figure 5)

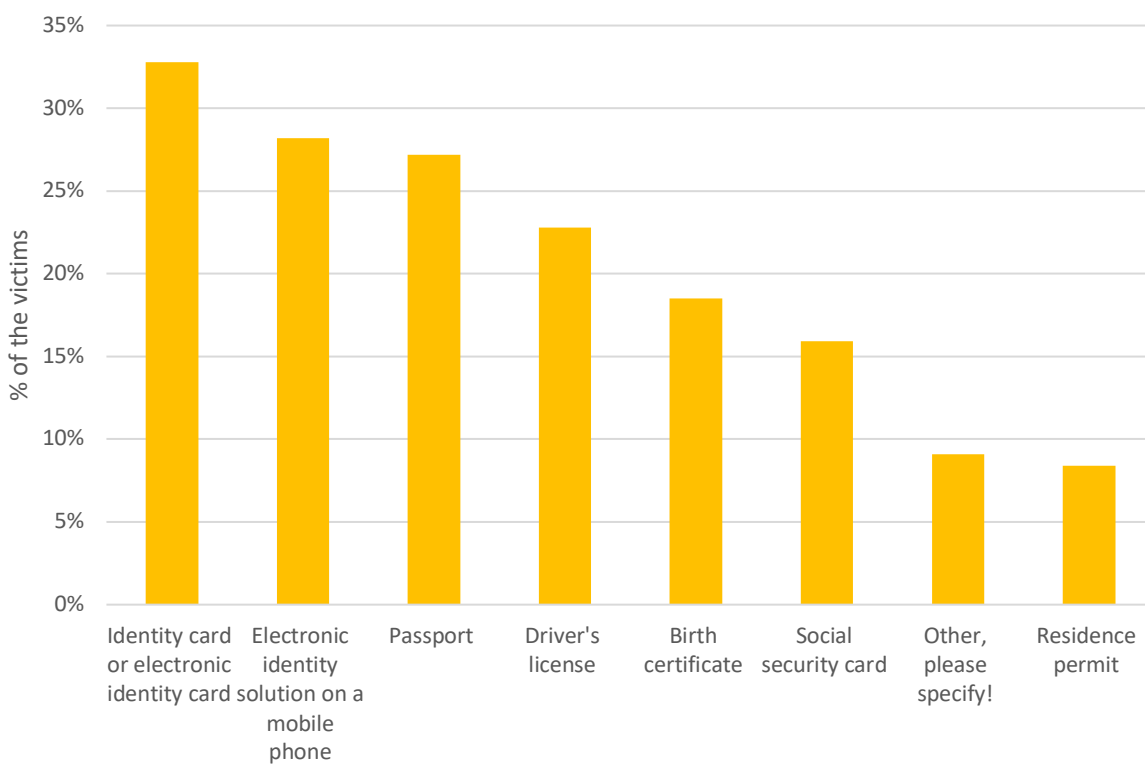

*Figure 5 – Which of the following identity documents or electronic identity solutions did the person misuse?*

Identity cards are, thus, highly popular identity documents that people carry with them on daily basis that tend to be also misused more frequently than other government issued identity solutions. In 68% of the cases physical card was presented (in an attempt) to impersonate the victim; 32% of the misuses of electronic identity cards involved electronic functionality of the card. Electronic identity solutions on mobile phones are also under widespread attacks.

---

[1] Earlier research has shown that over half of Europeans have experienced at least one type of fraud in the past two years (Finaso 2021), and 33% of the Americans have been the victim of identity theft at some point in their life (Fortunly 2022). The exact figures would vary significantly from one study to another depending on the exact definition of the fraud covered, etc.

Furthermore, 4% of respondents have experienced the misuse or attempted misuse of personal information, which involved applying for a new identity document, applying for government benefits or something else. For 44% of the victims, this involved (an attempt to) obtain a new identity card, for 33% a new electronic identity solution, for 20% a new passport and for 18% a new driver's license on their name. On several occasions, the frauds applied for multiple documents at once.

Notably, statistical analysis reveals that the personal experience of (attempted) misuse of identity documents does not make respondents less confident on the security of their government-issued identity documents.

The misuse and attempted misuse of financial accounts is about as widespread as the misuse of identity documents. 18% of the respondents who have a credit card have experienced (an attempted) misuse of their credit card, and 14% of the respondents have suffered from (an attempted) misuse of their bank accounts or other financial accounts. Notably, 36% of the victims are not sure how their financial accounts were accessed. The internet service or mobile app of the bank was misused in 30% of the cases and the fraud visited a bank office in person in 25% of cases. The later would involve obviously the use of a fraudulent identify document or the misuse of a genuine document belonging to the victim.

### 4.2.2. Which on-line solutions are at greater risk?

The various other online services are also under continued attacks, phishing e-mails being one of the most widespread means for obtaining the credentials of someone else's accounts. During the last 36 months, 45% of the respondents have received any fraudulent e-mails or visited fake websites that are designed for a person to reveal his or her username and password to hackers. 15% of the respondents indicate that they do not know, if they have experienced any *phishing* attempts, and 40% are confident that they have not experienced such attacks. We are rather sceptical about the later figure, and suspect that a notable share of potentially successful attacks that have went unnoticed.

Leaving the phishing attacks aside, Internet remains a highly contested space, as 25% of the respondents have experienced misuse or attempted misuse of their online accounts during the last 36 months. Social media accounts were involved for 50%, private e-mail for 42% and online shopping accounts for 34% of victims. (Figure 6)

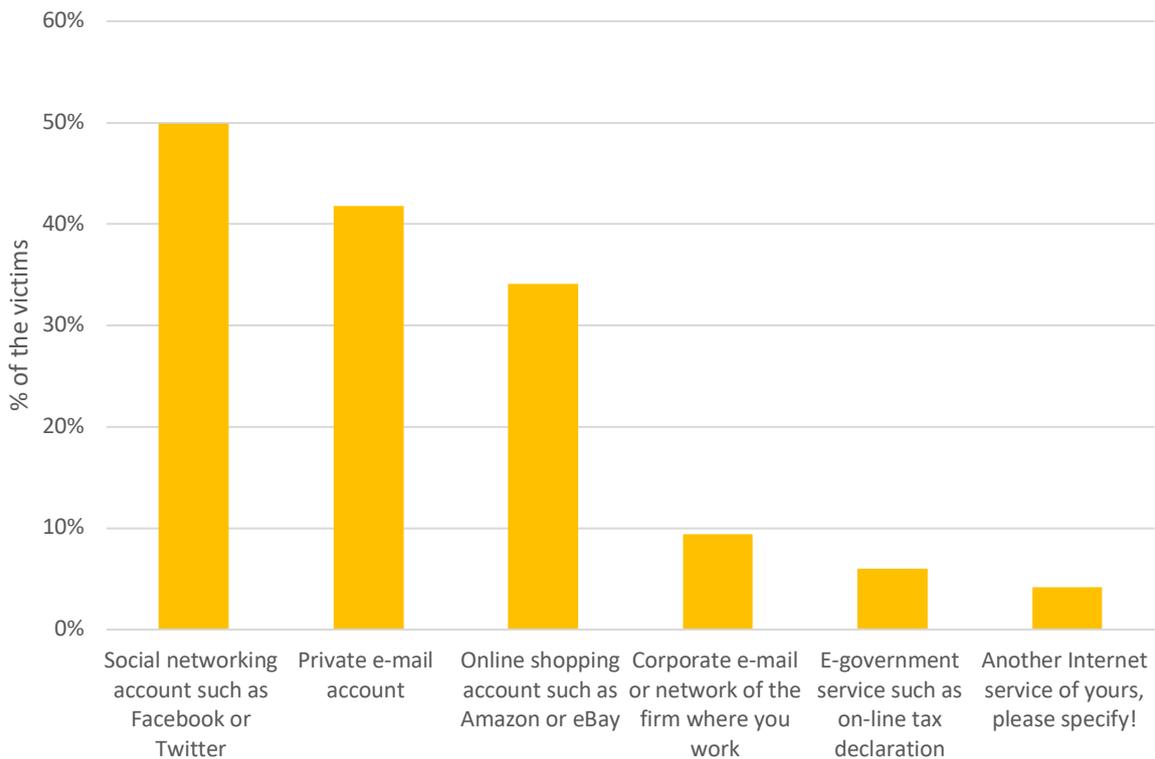

*Figure 6 – Which of the following types of your Internet accounts did the person use, or attempt to use, without your permission?*

On a positive note, 79% of respondents state that they use two-step authentication to protect their web services. We would have assumed that the take-up of two-step authentication is significantly lower in Europe and in the United States. It is, still, encouraging to see that there is an increasing understanding that password alone is no more sufficient for protection of online accounts.

What is more, people step up defences as the consequence of an (attempted) misuse of online accounts. Logistic regression reveals that the odds of using two-factor authentication are 1.3 times greater for those who experienced online fraud, compared to the odds of those who did not. Likewise, the odds of using two-factor authentication are 1.5 times greater for those who experienced phishing emails, compared to those who did not. Eventually, perhaps because of protective measures taken, victims of the misuse of online accounts do not appear to be less confident on the security of their online accounts than people with no direct experience of intrusion attempts.

### 4.2.3. The impact of identity theft

The misuse of attempted misuse of personal information occurred in 41% of the cases during the last 12 months, in 38% of cases more than a year, but less than two years ago. From the respondents, who have experienced the misuse or attempted misuse during the last 12 months, 49% experienced this once, 22% twice, 10% three times. A sizeable 20% of the victims indicate that the (attempted) misuse has happened even more frequently during the last 12 month. We consider this either an indication of ongoing attacks or an earlier unresolved identity theft, which remains truly problematic.

Only half of the victims know or think that they know how their personal information was taken. 30% of the respondents, who have experienced (an attempted) misuse of their personal information discovered the attempted misuse before my personal data was used. Half of the victims found out about it themselves, and the rest were notified by service providers. (Figure 7)

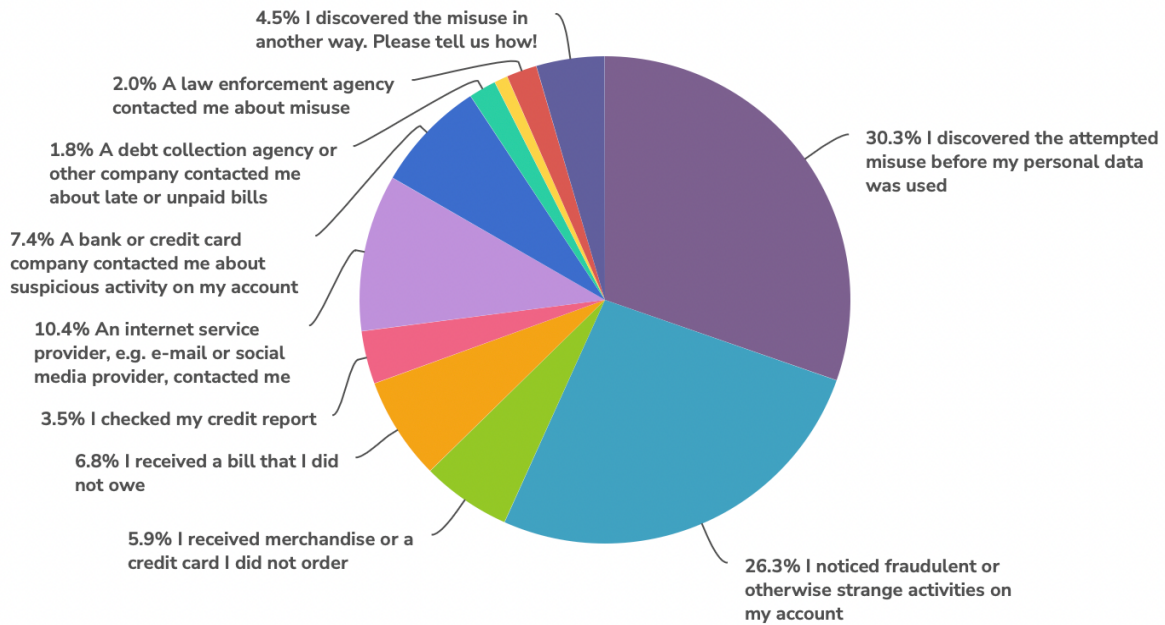

Figure 7 – How did you first find out someone had misused or attempted to misuse your personal information?

- 4.5% I discovered the misuse in another way. Please tell us how!
- 2.0% A law enforcement agency contacted me about misuse
- 1.8% A debt collection agency or other company contacted me about late or unpaid bills
- 7.4% A bank or credit card company contacted me about suspicious activity on my account
- 10.4% An internet service provider, e.g. e-mail or social media provider, contacted me
- 3.5% I checked my credit report
- 6.8% I received a bill that I did not owe
- 5.9% I received merchandise or a credit card I did not order
- 30.3% I discovered the attempted misuse before my personal data was used
- 26.3% I noticed fraudulent or otherwise strange activities on my account

35% of the victims of identity theft assess that their personal information was misused for one day or less before they discovered it. For 23% of the victims, it took up to more than a day, but less than a week to discover the misuse. For 15% of the victims, it took more than a week, but less than a month to discover the issue.

28% of the victims report that the thief was able to obtain financial benefits in the form of money, goods or services or something else, from the misuse of personal information. Notably, 16% of the victims do not know, if the thief was able to obtain financial benefits. No financial benefits were obtained in the remaining 55% of cases.

Only 37% of the respondents reported the misuse of personal information to the police. Yet, 64% of the persons, who reported the incident to the police were very satisfied or somewhat satisfied with the response of the police. From the persons, who did not report identity crime to the law enforcement, 26% thought that the police would not do anything, 25% did not find it a big issue to take the time to report it, and 17% did not know that they could report it. In 14% of the cases, the involved company, e.g., bank or Internet service provider reported the incident.

## 4.3. Societal acceptability of modern identity solutions
### 4.3.1. Government processing of identity data

The overall perception of complex technologies or products is an important determinant of the societal acceptability of such artefacts. 70% of the respondents believe that modern passports and electronic identity cards that include photos and fingerprints of document holders improve accuracy and convenience of identity checks, and protection from identity theft. 20% of the respondents are undecided, and less than 10% of the respondents disagree with the statement.

The understanding of the societal risks that modern identity documents may or may not bring about remains, however, very limited among respondents. Roughly 1/3 of the respondents believe that the modern identity documents increase the risk of hidden surveillance, misuse of personal information, intervention of their freedoms or unequal treatment of persons; another 1/3 of the respondents disagree with such view, and 1/3 remain undecided on this topic. The obvious need that what we see here is, therefore, that both the benefits and the potential risks of the biometric identity documents or electronic identity solutions need to be explained very carefully to the citizens to avoid unfounded opposition. (Figure 8)

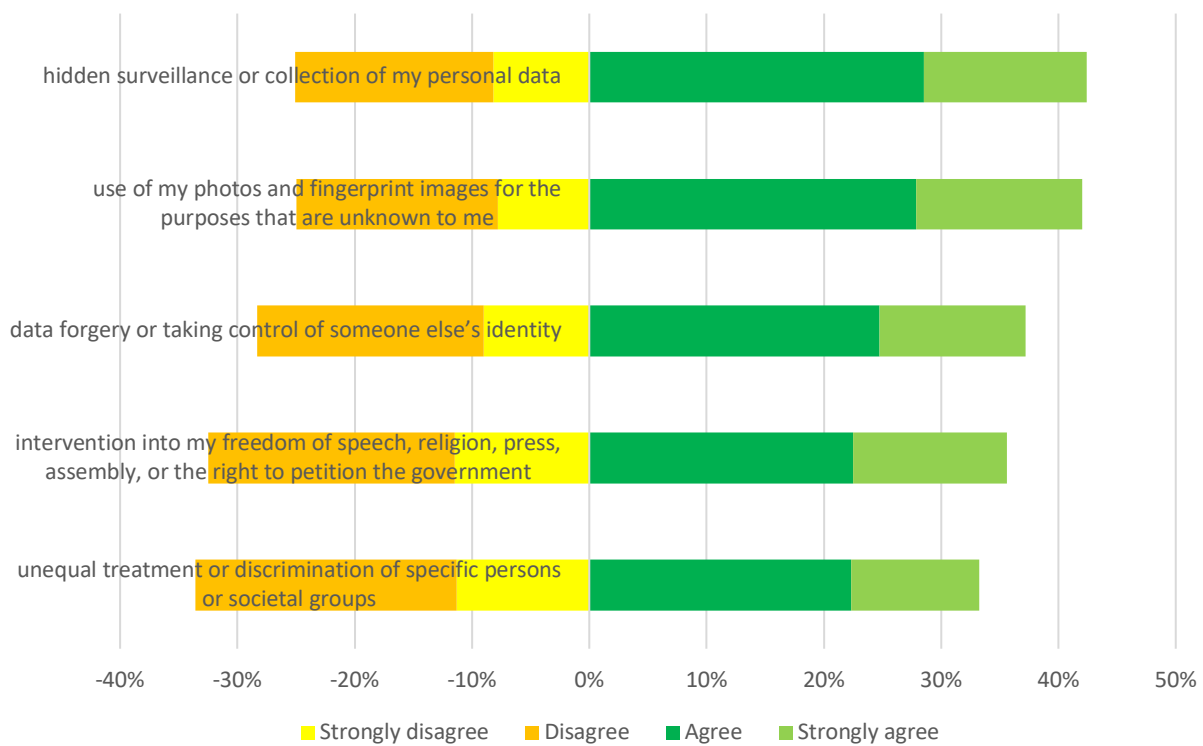

*Figure 8 – I believe that the modern passports and electronic identity cards that include photos and fingerprints of document holders increase the risk of*

*Note: Undecided respondents are excluded from this figure. The sum is therefore less than 100%.*

We find that the approval to government collecting and analysing the various types of information to provide secure means of identification, such as identity documents and electronic identity solutions, varies significantly by the type of data. 66% of the respondents agree with the government processing home addresses and photos that have been provided earlier to the authorities when applying for identity documents. There is also a significant support for government collecting and analysing fingerprint images (58%), personal identity codes (57%) and photos that have been provided earlier to the authorities (55%). Notably, the respondents oppose to the collection and analysis of geolocation data of the mobile phones and the photos that citizens have themselves published on Internet, e.g., on Facebook or LinkedIn. About 25% of the respondents remain undecided. Respondents find it especially difficult to judge, if they would agree with the eye iris images (29%), personal identity codes (28%) and DNA data (28%). (Figure 9)

*Figure 9 – I find it acceptable that my government collects and analyses the following types of information in order to provide secure means of identification, including identity documents and electronic identity*

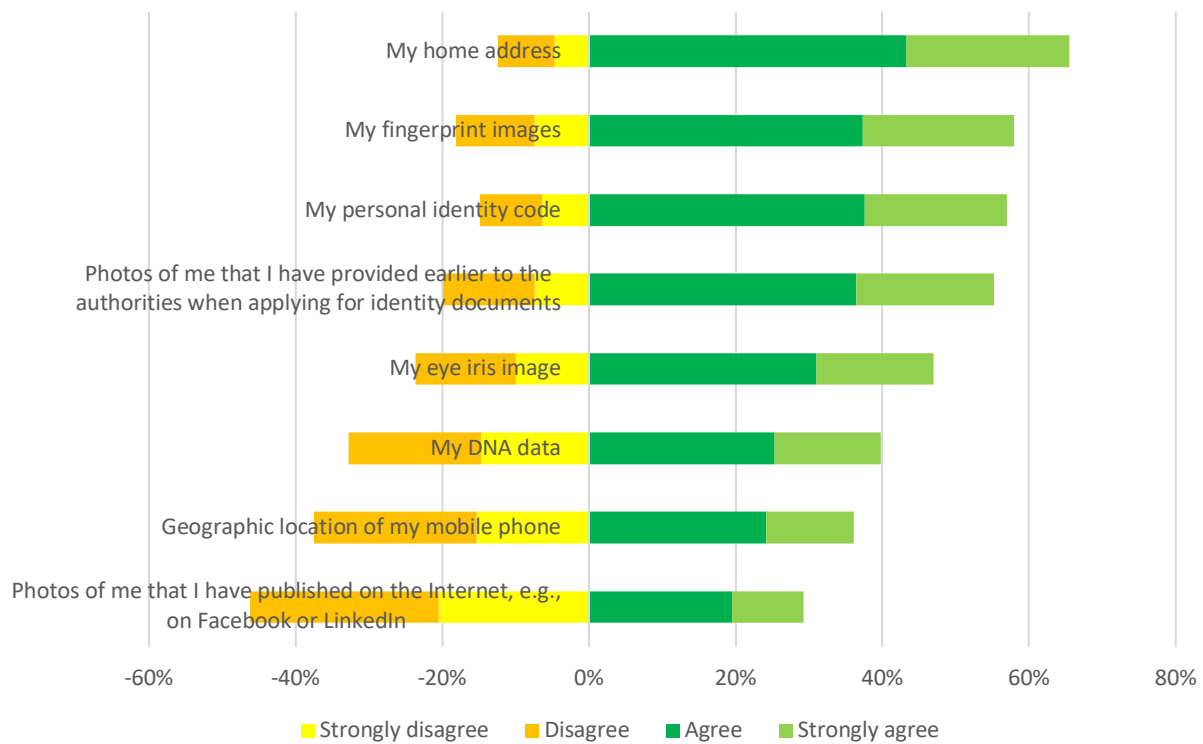

*Note: Undecided respondents are excluded from this figure. The sum is therefore less than 100%.*

A similar pattern emerges, when keeping the data from all passports and identity cards it has issued in a unified national registry. About 60% of the respondents agree with their names, photos and fingerprint images being stored in a unified national registry. Similarly to the previous question, about 25% of the respondents remain undecided, if they agree with their data being stored in a unified national registry. The undecidedness is the highest (29%) for the storage of personal identity codes.

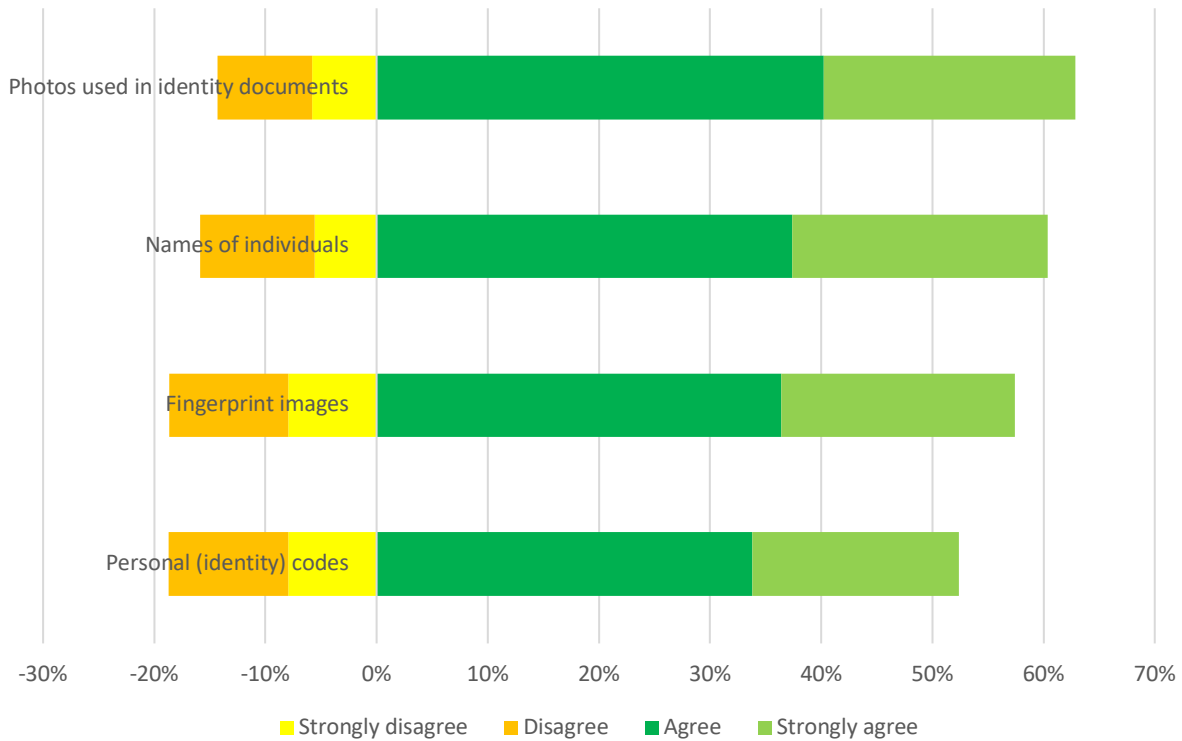

*Figure 10 – I find it acceptable that my government keeps the following data from all passports and identity cards it has issued in a unified national registry*

*Note: Undecided respondents are excluded from this figure. The sum is therefore less than 100%.*

Notably, 20-30% of the respondents remain undecided in the above questions whether they find such government data processing agreeable or not. This is most likely due to complicated technical nature of questions, whereas the broader implications of the choices concerned remain difficult to fully grasp for non-experts.

### 4.3.2. Issuance and renewal of identity documents

The different modalities for the submission of the document photos hold a largely similar level of support, when it comes to applying for a new government issued identity document. Only having a photo to be taken by a professional photographer stands out as a less agreeable modality. One might have guessed that younger people might be more technology savvy. Yet, we do not find any statistically significant differences in terms of age, gender or education level in terms of preferred means for submission of document photos.

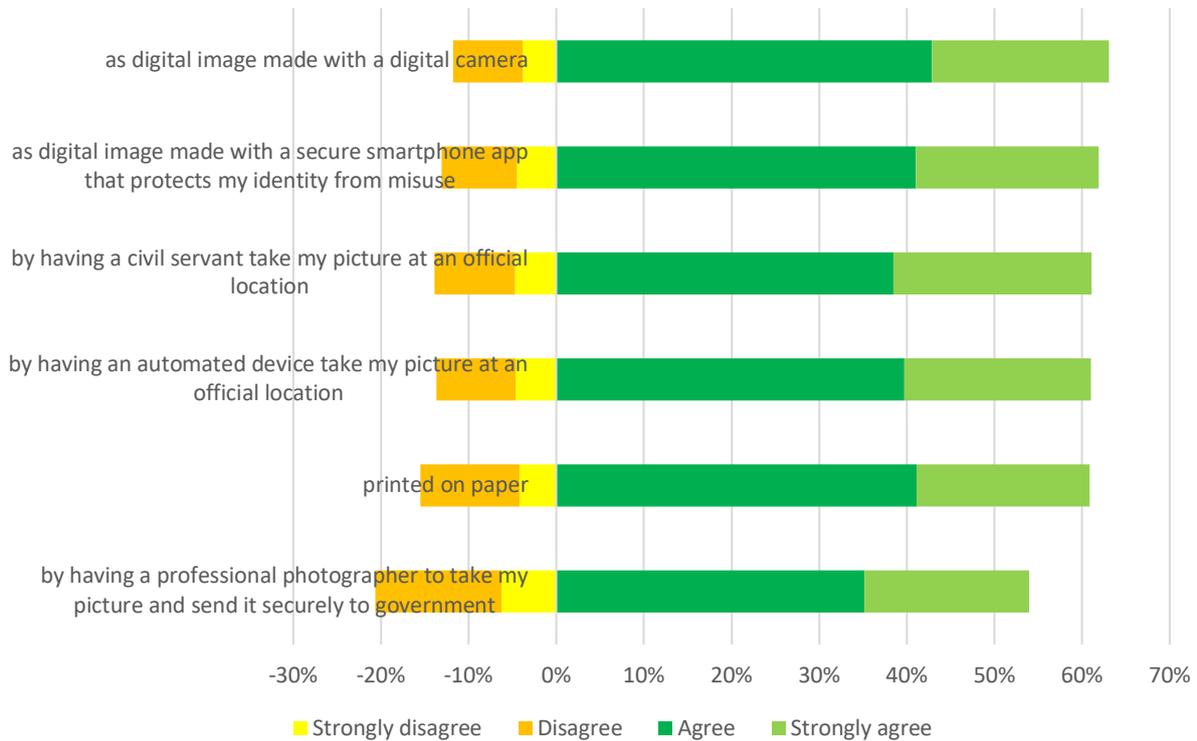

*Figure 11 – When applying for government issued identity document,
I could be ready to provide my photo:*

*Note: Undecided respondents are excluded from this figure. The sum is therefore less than 100%.*

The respondents, who would want to provide a printed photo, find that they can choose the best photo of themselves this way. Half of the respondents believe that printed photos are more tamper proof than digital files. Yet, there is also a significant share of the respondents, who are ready to provide a photo – in an accustomed manner – on paper even if they find printed photos less convenient to handle than digital files.

We find that there is broadly based 70% support to manual processing of the photos by civil servants, when it comes to comparing the photo that submitted for the renewal of my passport or identity card with the photo in my previous identity document. Contrastingly, only 53% of the respondents agree with the automated data processing by with artificial intelligence. Germany and the United States stand out in comparison to other countries with a significantly lower support to such use of AI. This is both due to greater share of undecided responders and a smaller share of the responders who agree with such use of artificial intelligence. (Figure 12) The above findings are most likely to hold not only for renewal of identity documents, but also for other 1:1 identity checks.

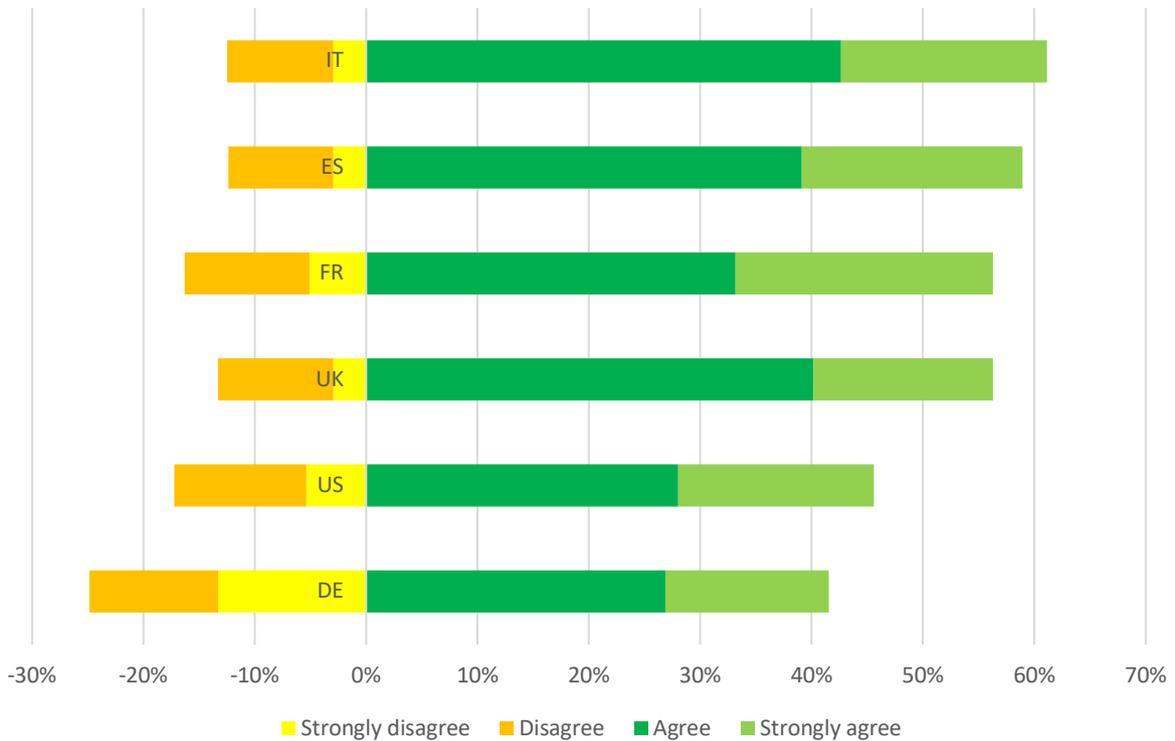

*Figure 12 – I find it acceptable that the photo I submit for the renewal of my passport or identity card is compared with the photo in my previous identity document automatically, by artificial intelligence*

*Note: Undecided respondents are excluded from this figure. The sum is therefore less than 100%.*

65% of the respondents find it important that there is human oversight of automated decisions made by artificial intelligence, when it comes to in processing of my identity document photos and fingerprints by the government. Conversely, 55% of the respondents find it important that there is an automated oversight of human-made decisions. The agreeability of the use of artificial intelligence in identity management and the organisation of oversight remains clearly challenging issue for many respondents to take an informed view. 30% of the respondents are undecided on automated oversight with artificial intelligence, and 25% do not have a clear view, if human oversight of the automated decisions made by artificial intelligence would be necessary.

Similarly to the previous question on document photos, we do not find any statistically significant differences in terms of age, gender or education level when it comes to the use of artificial intelligence in processing document photos; human or automated oversight of the image processing.

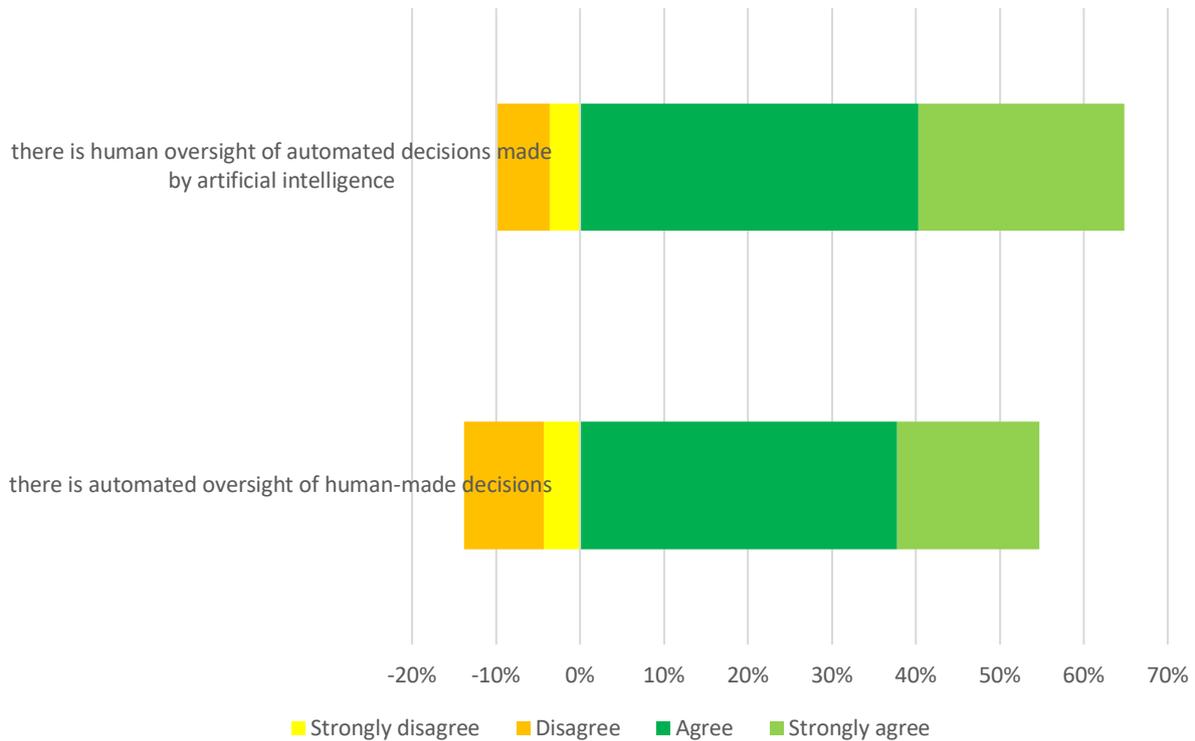

*Figure 13 – I find it important that in processing of my identity document photos and fingerprints by the government:*

*Note: Undecided respondents are excluded from this figure. The sum is therefore less than 100%.*

### 4.3.3. Identity checks and electronic identity wallets

An increasing number of countries consider the introduction of novel digital identity wallets. One can think of digital identity wallets as a secure smartphone app that could use instead of driver's license, identity card or passport to prove your identity. Such digital identity wallets would include document holder's facial image, biographical data, and would be protected with digital security features.

We find that there is there is a broad interest in using the various functionalities of the digital identity wallet. More than half of the respondents are interested in using one or another functionality of the wallet, while about 25-30% of the respondents are undecided, and a smaller group of respondents are uninterested in such novel tool.

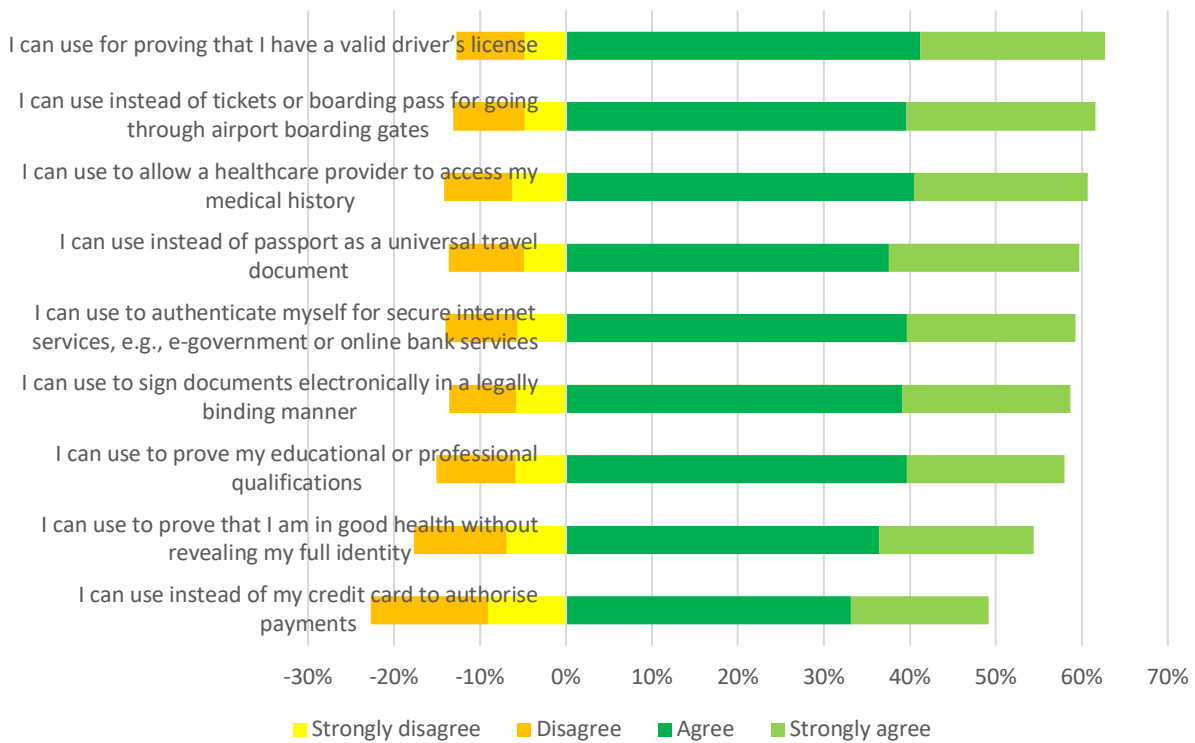

*Figure 14 – I am interested in using a digital identity wallet, that…*

*Note: Undecided respondents are excluded from this figure. The sum is therefore less than 100%.*

Once again, the age, gender and education level of the respondents does not have any significant explanatory power when it comes to interest in using a digital ID wallet. 70% of the respondents who have experienced (an attempted) misuse of identity documents would be interested in using a digital identity wallet as an universal travel document. In comparison, only 58% of the respondents with no direct experience of identity document fraud would be interested in using a digital identity wallet. Likewise, personal experience with identity document fraud increases the interest in using a digital identity wallet instead of driver's licence (73% of respondents), for secure online authentication (78%), or signing documents electronically (69%). However, a further statistical analysis revealed that the personal experience with misuse of identity documents has a statistically very week explanatory power in explaining the interest in using the various functionalities of the electronic identity wallet.

### 4.3.4. Government data sharing

More than half of the respondents agree with their government sharing their identity document data with relevant government agencies domestically, and with their government obtaining the identity document data foreigners from their home countries. There is, however, less support to the government sharing identity document data with the governments of the European Union member states. In Spain, France and Italy, 55-60% of the respondents agree with their identity document data being shared with other governments in the European Union; in Germany, United Kingdom and the United States around 44% of the respondents agree with such data sharing. However, age, gender, education level nor the experience of identity document fraud do not have a statistically significant explanatory power with a view to agreeability of the government information sharing.

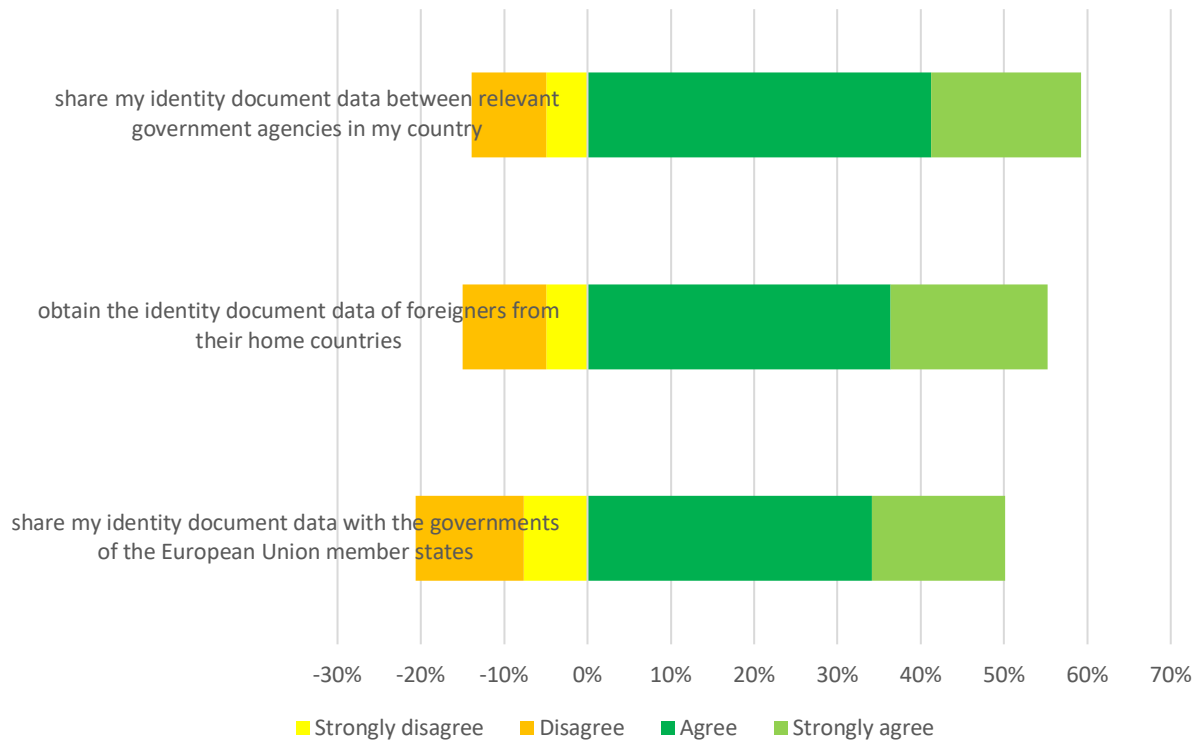

*Figure 15 – I find it acceptable that my government would do the following to facilitate an official investigation…*

*Note: Undecided respondents are excluded from this figure. The sum is therefore less than 100%.*

Majority of the respondents agree also with the government sharing identity document data with private entities. However, the disapproval of such data sharing is notably stronger than the opposition against data sharing within and between the government(s). As an example, 30% of the respondents disagree with data sharing with banks and credit card companies. In Germany 39% and in France 36% of the respondents disagree with their identity document data being shared with banks. The United States, United Kingdom and Spain, only 23-25% of the respondents the idea of such data sharing.

The victims of the misuse of bank or other financial accounts are, perhaps somewhat counterintuitively, more likely to agree with government sharing their identity document data with banks.

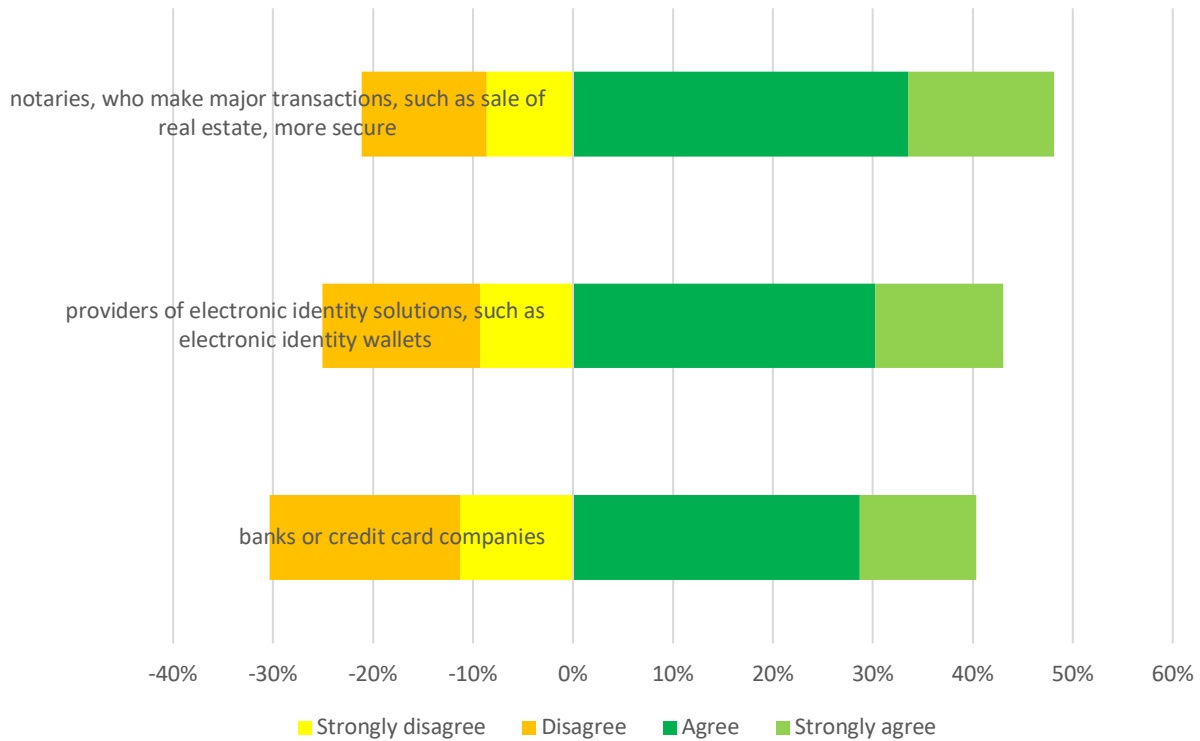

*Figure 16 – I find it acceptable that my government would share my identity document data with private entities, such as:*

Note: Undecided respondents are excluded from this figure. The sum is therefore less than 100%.

## 5. Discussion and conclusions

In the following, we discuss the results of the current societal acceptability study and draw recommendations that facilitate the adoption of modern identity solutions.

**Confidence in public and private sector identity solutions**
As pointed out above, government-issued identity documents enjoy broad confidence that they are secure. However, the general public places significantly less trust in the security of various private sector services. Notably, one-third of the respondents are unconfident or very unconfident that their social media accounts, such as Facebook or Twitter, are secure. This is against the background that half of the respondents have experienced (an attempted) misuse of their social media accounts. Additionally, personal email accounts are reported to be under widespread attacks. This is especially significant when considering the widespread use of social media and personal email accounts to facilitate password-less access to various other online services. Clearly, bolstering the security of social media and personal email accounts with strong two-step authentication remains a major issue to be addressed. This is where a globally standardised identity token, such as digital identity wallet could potentially make a significant contribution.

**Issuance and renewal of identity documents**
While government issued identity documents enjoy high levels of confidence that they are secure, a remarkable 16% of the respondents state that someone used or attempted to use one of their identity documents or electronic identity solutions, without their permission, during the past 36 months. However, respondents who experienced misuse of a type of document failed to have lower levels of confidence in that same type of document. Respondents of the survey believe that modern passports and electronic identity cards increase protection from identity theft and improve the accuracy and convenience of identity checks. This, along with trust in the security of identity documents, establishes a solid basis for further improvement the identity documents and related back-office processes.

About 60% of the respondents agree with government collecting and analysing home addresses, fingerprint images, personal identity codes and photos of the citizens, which have been provided earlier to the authorities, to provide secure means of identification. The respondents in different countries are generally favourable regarding the unified national registry of photos used in identity documents, and regarding the processing of the photos that have been provided earlier to the authorities.

People are very open for different modalities of submitting a document photo, be it a printed photo or a taking a digital image with a secure smartphone app. The above opens potentially also avenues for remote identity onboarding as an emerging field in identity management (Laas-Mikko et al. 2022). However, the use of the photos published by the citizens themselves online, e.g. in social media, should be avoided in all stages of identity document life cycle.

**Identity checks and verification of identity documents**
The agreeability of use of artificial intelligence in checking the identity of persons remains a challenging topic as 30% of the respondents remain undecided on desirability of the use of such technologies. This is most likely due to the novelty of the topic and insufficient understanding of the potential consequences of such use technology. International media coverage that has highted the societal risks of the use of facial recognition technologies identification of persons for covert surveillance or 'social scoring' purposes is likely also a factor. The risks that are involved in using artificial intelligence for verification identity and the related mitigation measures are, therefore, to be carefully explained to the citizens before actual adoption of new technologies.

We take also note that there is a strong interest in using digital identity wallets. There is a broad interest in the various functionalities that digital identity wallets can offer, such as proving that one has a valid driver's licence, replacing passports as universal travel documents, proving one's identity online, or signing documents electronically. We take also a note that a relatively high share of the respondents are undecided on the benefits, people are also undecided on the risks and on specific use cases. We would consider this in the context general trust towards government while lacking technical knowledge for independent assessment. Hence, there seems to be room for activities to increase public awareness.

Although some country-specific differences exist, it is generally acceptable that artificial intelligence is used to compare the photo in one's previous identity documents; preferably with a human oversight of automated decisions made by the artificial intelligence. The above findings are largely in-line with earlier research that has demonstrated that people are more likely to put less trust on artificial intelligence than humans, and accept the mistakes made by humans more readily (Hidalgo et al. 2021).

**Government data sharing**
60% of the respondents of the survey find it acceptable that the government shares, as a part of official investigation, identity document data between relevant government agencies; 50% would accept their identity document data being shared with other governments in the European Union. Spain, France and Italy are more open to such data sharing with the EU governments, while Germany, the United Kingdom and the United States are more reluctant to do so.

There is less of support for government sharing identity document data with private entities, in comparison to government agencies. Yet, careful explanation of purpose of the reuse of the identity data that the government has collected and of the related safeguards is likely to change the views in this question. We find as well, that victims of bank account misuse or other types of financial account misuse are significantly more likely to support their government data being sharing with banks.

In conclusion, we find that there is a sufficient public support to the adoption of the technologies that are being developed in iMARS. However, the purpose and the intended use of any personal information, such as facial images, has to be carefully explained to the public; and fallback procedures need to be put in place to correct for possible mistakes in automated decision making. Also, independent oversight of the actual use of such data must be ensured to bolster privacy protection and to avoid any discrimination.

**Concluding remarks**
Earlier research (e.g., Ng-Kruelle et al. 2006, Tiits et al. 2014a, Tiits et al. 2014b) has shown that differences exist in between the individual EU countries and the US in their identity management systems and on the public opinion. Based on the results of the current study, however, we can observe significant convergence of the public opinion in the countries analysed[2]. There are no significant statistical differences by country other than respondents from Germany are slightly more concerned about their privacy, and are more sceptical about biometrics, centralised identity document databases, etc. This is aligned with the previous studies, although the differences have closed. The likely reason for an observed convergence of views is that the identity management systems of various countries and people's experience are becoming more similar. What is more, a significant share of respondents needs further information to be able to formulate their views on agreeability of the use of biometric data and advanced technologies, such as artificial intelligence, in identity checks.

Our research has certain limitations. One key limitation of our study lies in its scope of data collection, which is restricted to six countries in the European Union and the United States.

---

[2] A two-step approach was undertaken: Regressing dummy variables onto each variable mentioned in various survey questions and looking for relationships that were a) statistically significant and b) explained some appreciable amount of variation.

As such, the generalisability of our findings to other cultural and geopolitical contexts remains questionable. Variations in cultural norms, legal frameworks, digital literacy levels, and experiences with identity theft could significantly impact the societal acceptability of modern identity solutions. Consequently, further research should aim to expand the geographical scope to include a broader array of countries.

Another limitation pertains to the cross-sectional nature of the study, which only provides a snapshot of public opinion at one point in time. As technology continues to advance rapidly and societal attitudes evolve, it would be beneficial for future research to adopt a longitudinal design, tracking changes in attitudes over time. This would offer a more dynamic understanding of the factors influencing societal acceptability, helping to identify trends and predict future behaviour.

Moreover, while the aggregate respondent pool is sufficiently large for general conclusions, deeper examinations based on demographic parameters may be unreliable due to the limited representation of certain subsets (for instance, specific age brackets, occupational categories, and so on). Consequently, future studies would likely benefit from integrating qualitative methodologies such as comprehensive interviews, focus group discussions, or case analyses. These approaches could offer even a better understanding of the motivations underpinning the acceptance or rejection of contemporary identity solutions by various individuals.

In terms of specific further research directions, our study highlights the need to understand the public's view of AI in identity management more thoroughly. We found that 30% of the respondents remained undecided about the use of AI in identity checks. A deeper investigation into the reasons behind this indecision could prove beneficial in mitigating concerns and guiding policy decisions. Given the increasing role of AI in many sectors, a dedicated study focusing on AI in identity management would be timely and could greatly inform both technology developers and policymakers.

More research is also needed on the acceptability of government data sharing with private entities. This is a particularly sensitive and complex issue with implications for privacy, trust, and security. Understanding the conditions under which the public might find such sharing acceptable, and the measures that can be put in place to ensure security and privacy, could offer valuable insights for policy and practice.

Finally, despite some critiques pertaining to the simplistic and unadorned application of technology acceptance models, most notably those voiced by Bannister (2023), our research found these models to remain fundamentally pertinent and applicable when examining the societal acceptance of electronic identity solutions. Considering the policy implications underscored by our study, the ability to ground our research within an academic framework proved beneficial. This not only facilitated the research process but also ensured a translation of our findings in a manner that resonated with policy discourse, consequently enhancing the policy relevance of our work.

# Acknowledgements

This project has received funding from the European Union's Horizon 2020 research and innovation programme under grant agreement No 883356.

# Ethics disclaimer

The international survey that serves as the foundation for this paper involved human participants. Adherence to the requirements of the GDPR was carefully ensured during data collection. Prior to gathering data, participants provided informed consent for the collection and processing of their information. No personally identifying information, nor details regarding political or sexual preferences, was gathered from participants. Consequently, no ethics committee approval was required for this research.

# References

Ajzen, I. (1985) 'From intentions to actions: A theory of planned behavior.' In J. Kuhl & J. Beckmann (Eds.), Action Control: From Cognition to Behavior, New York: Springer-Verlag. 3, pp. 11-39.

Akdemir, N. (2021). Coping with Identity Theft and Fear of Identity Theft in the Digital Age. In: López Rodríguez, A.M., Green, M.D., Kubica, M.L. (Eds) Legal Challenges in the New Digital Age, 176-197. Leiden: Koninklijke Brill NV.

Alchemer (2023). Enterprise online software & tools, https://alchemer.com

Apple (2023). App Store stopped more than $2 billion in fraudulent transactions in 2022, Newsroom, 16 May 2023, https://apple.co/3BDirms.

Ash, J. (1997). Factors for Information Technology Innovation Diffusion and Infusion in Health Sciences Organizations: A Systems Approach, Portland State University, 1997.

Bannister, F. (2023), "Beyond the box: Reflections on the need for more blue sky thinking in research", *Government Information Quarterly*, Vol. 40 No. 3, p. 101831, doi: 10.1016/j.giq.2023.101831.

Betz-Hamilton, A. (2022), A Comparison of the Financial, Emotional, and Physical Consequences of Identity Theft Victimization Among Familial and Non-Familial Victims, Journal of Financial Counseling and Planning, 33(2), 217–227, doi: 10.1891/JFCP-2021-0014.

Boneh, D., Grotto, A. J., McDaniel, P., Papernot, N. (2019). How Relevant Is the Turing Test in the Age of Sophisbots? IEEE Security & Privacy, 17(6), 64-71.

Brandtz, P.B., J. Heim & A. Karahasanovi (2011). Understanding the new digital divide-A typology of Internet users in Europe. International Journal of Human-Computer Studies, 69(3), 123-138.

Cint (2023). Digital insights gathering platform, https://cint.com

Dang, H., Liu, F., Stehouwer, J., Liu, X., Jain, A.K. (2020). On the Detection of Digital Face Manipulation. 2020 IEEE/CVF Conference on Computer Vision and Pattern Recognition (CVPR), 5780-5789.

Davis, F.D. (1989). "Perceived usefulness, perceived ease of use, and user acceptance of information technology", MIS Quarterly, 13(3), 319–340.

Eurostat (2022). Eurostat database, Eurostat, https://ec.europa.eu/eurostat/data/database.

Finaso (2021). One in Five Europeans Have Experienced Identity Theft Fraud in the Last Two Years, https://bit.ly/3Eo6q4V.

Fishbein, M. & Ajzen, I. (1975) Belief, attitude, intention and behavior: An introduction to theory and research, MA: Addison-Wesley.

Fortunly (2022). 20 Worrying Identity Theft Statistics for 2022, https://bit.ly/3uRTnpe.

Harrell, E. (2014). Victims of identity theft. Bureau of Justice Statistics. https://www.bjs.gov/content/pub/pdf/vit14.pdf.

Hidalgo, C. A., Orghiain, D., Canals, J. A., de Almeida, F., & Martin, N. (2021). How Humans Judge Machines. In How Humans Judge Machines. The MIT Press. https://doi.org/10.7551/MITPRESS/13373.001.0001

Javelin (2020). Identity Fraud Study: Genesis of the Identity Fraud Crisis. https://bit.ly/3rFrrmi.

Kalvet, T., Tiits, M., Ubakivi-Hadachi, P. (2019a). Risks and Societal Implications of Identity Theft. In: Chugunov, A., Misnikov, Y., Roshchin, E., Trutnev, D. (Eds). Electronic Governance and Open Society: Challenges in Eurasia: 5th International Conference, EGOSE 2018, St. Petersburg, Russia, November 14-16, 2018, Revised Selected Papers. Springer.

Kalvet, T., Tiits, M., Laas-Mikko, K. (2019b). Public Acceptance of Advanced Identity Documents. In: Ojo, A.; Kankanhalli, A.; Soares, D. (Ed.). Proceedings of the 11th International Conference on Theory and Practice of Electronic Governance, 429–432. Galway, Ireland. https://doi.org/10.1145/3209415.3209456.

Kantar (2020). Europeans' attitudes towards cyber security. Special Eurobarometer 499, https://bit.ly/3xAwicq.

Kemppainen, L., Kemppainen, T., Kouvonen, A., Shin, Y.-K., Lilja, E., Vehko, T. and Kuusio, H. (2023), "Electronic identification (e-ID) as a socio-technical system moderating migrants' access to essential public services – The case of Finland", Government Information Quarterly, p. 101839, doi: 10.1016/j.giq.2023.101839.

Kindt, E., Lopez, C.A.F., Tiits, M., Kalvet, T. (2021). Legal, ethical and societal requirements. Deliverable 3.1, iMARS, https://bit.ly/3KYZkpO.

King, W.R. and He, J. (2006). "A meta-analysis of the technology acceptance model", Information & Management, 43(6), 740–755.

Laas-Mikko, K., Kalvet, T., Derevski, R., Tiits, M. (2022). Promises, Social, and Ethical Challenges with Biometrics in Remote Identity Onboarding. In: Rathgeb, C., Tolosana, R., Vera-Rodriguez, R., Busch, C. (eds) Handbook of Digital Face Manipulation and Detection. Advances in Computer Vision and Pattern Recognition. Springer, Cham. https://doi.org/10.1007/978-3-030-87664-7_20.

Mathieson, K. (1991). "Predicting User Intentions: Comparing the Technology Acceptance Model with the Theory of Planned Behavior", Information Systems Research, 2(3), 173–191.

Ng-Kruelle, G., Swatman, P. A., Hampe, J. F., & Rebne, D. S. (2006). Biometrics and e-Identity (e-Passport) in the European Union: End-User Perspectives on the Adoption of a Controversial Innovation. Journal of Theoretical and Applied Electronic Commerce Research, 1(2), 12–35. https://doi.org/10.3390/JTAER1020010

Oudekerk, B., Langton, L., Warnken, H., Greathouse, S.M., Lim, N., Taylor, B., Welch, V. (2018). Building a National Data Collection on Victim Service Providers: A Pilot Test. Bureau of Justice Statistics. https://www.ncjrs.gov/pdffiles1/bjs/grants/251524.pdf.

Reyns, B.W. Identity-Related Crimes. In: Reichel, R.; Randa, R. (eds.) Transnational Crime and Global Security, 161-179- Praeger Security International, 2018.

Rogers, E.M. (2003) Diffusion of innovations, New York: Free Press.

Taherdoost, H. (2018), "A review of technology acceptance and adoption models and theories", Procedia Manufacturing, Vol. 22, pp. 960–967, doi: 10.1016/j.promfg.2018.03.137.

Tamilmani, K., Rana, N.P., Wamba, S.F. and Dwivedi, R. (2021), "The extended Unified Theory of Acceptance and Use of Technology (UTAUT2): A systematic literature review and theory evaluation", International Journal of Information Management, Vol. 57, p. 102269, doi: 10.1016/j.ijinfomgt.2020.102269.

Tiits, M., Kalvet, T., Laas-Mikko, K. (2014a). Analysis of the ePassport readiness in the EU. FIDELITY Deliverable 2.2. Tartu: Institute of Baltic Studies.

Tiits, M., Kalvet, T., Laas-Mikko, K. (2014b). Social Acceptance of ePassports. In: Brömme, A.; Busch, C. (eds.) Proceedings of the 13th International Conference of the Biometric Special Interest Group. IEEE, Darmstadt.

Tiits, M., Ubakivi-Hadachi, P. (2015). Common use patterns of identity documents. EKSISTENZ D9.1. Tartu: Institute of Baltic Studies.

Tiits, M., Ubakivi-Hadachi, P. (2016). Societal risks deriving from identity theft. EKSISTENZ D9.2. Tartu: Institute of Baltic Studies.

Tiits, M., Kalvet, T. (2022). Identity theft and societal acceptability of electronic identity in Europe and in the United States: survey questionnaire, Tartu: Institute of Baltic Studies, https://doi.org/10.23657/imars-2022.

TNS Opinion & Social (2017). Europeans' attitudes towards cyber security. Special Eurobarometer 464a. https://bit.ly/3xKo5Cr.

Tolosana, R., Rathgeb, C., Vera-Rodriguez, R., Busch, C., Verdoliva, L., Lyu, S., Nguyen, H. H., Yamagishi, J., Echizen, I., Rot, P., Grm, K., Štruc, V., Dantcheva, A., Akhtar, Z., Romero-Tapiador, S., Fierrez, J., Morales, A., Ortega-Garcia, J., Kindt, E., Jasserand, C., Kalvet, T., Tiits, M. (2022). Future Trends in Digital Face Manipulation and Detection. In: Advances in Computer Vision and Pattern Recognition, 463–482. https://link.springer.com/chapter/10.1007/978-3-030-87664-7_21.

U.S. Government Accountability Office (2014). Identity theft: Additional Actions Could Help IRS Combat The Large, Evolving Threat of Refund Fraud. Report to Congressional Requesters, GAO-14-633. https://www.gao.gov/assets/670/665368.pdf.

Venkatesh, V. (2000). "Determinants of Perceived Ease of Use: Integrating Control, Intrinsic Motivation, and Emotion into the Technology Acceptance Model", Information Systems Research, 11(4), 342–365.

Venkatesh, V., Morris, M.G., Davis, G.B., Davis, F.D. (2003). "User acceptance of information technology: toward a unified view", MIS Q, 27(3), 425–478.

Venkatesh, V., Thong, J.Y.L. and Xu, X. (2012), "Consumer Acceptance and Use of Information Technology: Extending the Unified Theory of Acceptance and Use of Technology", MIS Quarterly, Management Information Systems Research Center, University of Minnesota, Vol. 36 No. 1, pp. 157–178, doi: 10.2307/41410412.

Zhong, Y., Oh, S., & Moon, H. C. (2021). "Service transformation under industry 4.0: Investigating acceptance of facial recognition payment through an extended technology acceptance model", Technology in Society, 64, 101515. https://doi.org/10.1016/J.TECHSOC.2020.101515